\documentclass[journal=jacsat,manuscript=article]{achemso}

\usepackage{chemformula}
\usepackage[T1]{fontenc}
\usepackage{siunitx}
\usepackage{soul}
\usepackage{dsfont}
\usepackage{soul}

\author{Jonathan R. Church}
\affiliation{J.R.C. and O.B. contributed equally to this work}
\alsoaffiliation{School of Chemistry, Tel Aviv University, Tel Aviv 6997801, Israel}
\author{Ofir Blumer}
\affiliation{J.R.C. and O.B. contributed equally to this work}
\alsoaffiliation{School of Chemistry, Tel Aviv University, Tel Aviv 6997801, Israel}
\author{Tommer D. Keidar}
\affiliation{School of Chemistry, Tel Aviv University, Tel Aviv 6997801, Israel}
\author{Leo Ploutno}
\affiliation{School of Chemistry, Tel Aviv University, Tel Aviv 6997801, Israel}
\author{Shlomi Reuveni}
\alsoaffiliation{School of Chemistry, Tel Aviv University, Tel Aviv 6997801, Israel}
\affiliation{The Center for Computational Molecular and Materials Science, Tel Aviv University, Tel Aviv 6997801, Israel}
\altaffiliation{The Center for Physics and Chemistry of Living Systems, Tel Aviv University, Tel Aviv 6997801, Israel}
\author{Barak Hirshberg}
\alsoaffiliation{School of Chemistry, Tel Aviv University, Tel Aviv 6997801, Israel}
\affiliation{The Center for Computational Molecular and Materials Science, Tel Aviv University, Tel Aviv 6997801, Israel}
\altaffiliation{The Center for Physics and Chemistry of Living Systems, Tel Aviv University, Tel Aviv 6997801, Israel}
\email{hirshb@tauex.tau.ac.il}

\title[title for paper]
  {Accelerating Molecular Dynamics through Informed Resetting}

\abbreviations{ISR,SR,MFPT,MD,MetaD}

\begin{document}

\begin{abstract}
We present a procedure for enhanced sampling of molecular dynamics simulations through informed stochastic resetting. Many phenomena, such as protein folding and crystal nucleation, occur over time scales inaccessible \textcolor{black}{in} standard simulations. We recently showed that stochastic resetting can accelerate molecular simulations that exhibit broad transition time distributions. However, standard stochastic resetting does not exploit any information about the reaction progress. \textcolor{black}{For a model system and chignolin in explicit water,} we demonstrate that an informed resetting protocol leads to greater accelerations than standard stochastic resetting \textcolor{black}{in} molecular dynamics and Metadynamics simulations. This is achieved by resetting only when a certain condition is met, e.g., when the distance from the target along the reaction coordinate is larger than some threshold. 
\textcolor{black}{We use these accelerated simulations to} infer \textcolor{black}{important kinetic observables such as} the unbiased mean \textcolor{black}{first-passage time} \textcolor{black}{and direct transit time. For the latter, Metadynamics with informed resetting leads to speedups of $2\text{-}3$ orders of magnitude over unbiased simulations with relative errors of only $\sim35\text{-}70\%$.} Our work significantly extends the applicability of stochastic resetting for enhanced sampling of molecular simulations.  
\end{abstract}

\section{Introduction}

Molecular dynamics (MD) is a powerful tool that is commonly employed to gain physical insights into complex chemical systems. Unfortunately, using standard MD to study processes which occur over time scales longer than a few microseconds, such as crystal nucleation and protein dynamics, is currently not feasible~\cite{Tiwary2013,Kleiman2023,Salvalaglio2014}. 
Over the years, a number of methods have been proposed to overcome this time scale issue, such as umbrella sampling\cite{Kastner2011,Torrie1977}, adiabatic free-energy dynamics\cite{Rosso2002,Rosso2002a}, Metadynamics (MetaD)\cite{Barducci2008,Barducci2011,Valsson2016,Sutto2012,Bussi2020}, on-the-fly probability enhanced sampling (OPES)\cite{Invernizzi2020,Invernizzi2021,Invernizzi2022,Invernizzi2020a}, Milestoning\cite{Faradjian2004, Elber2020}, and stochastic resetting for enhanced sampling~\cite{Blumer2022, Blumer2024}.

In this paper, we focus on stochastic resetting (SR) \cite{Pal2017, Chechkin2018, Evans2020,Kundu_2024}, in which trajectories are randomly stopped and re-initiated with independent and identically distributed samples of positions and momenta\cite{Blumer2022, Blumer2024}. It is a collective variables-free approach that can be employed as a stand-alone method\cite{Blumer2022} or combined with other enhanced sampling algorithms, such as Metadynamics.\cite{Blumer2024}. SR also minimally perturbs the natural dynamics of the system, therefore providing reliable estimates of the unbiased kinetics.\cite{Blumer2022, Blumer2024,10.1063/5.0243783}.  

One key advantage of SR is that it requires no prior knowledge to accelerate the simulations. However, \textcolor{black}{the probability of resetting is \textcolor{black}{then} agnostic to the reaction progress.} In other words, \textcolor{black}{when the resetting time comes}, the simulation will be reset no matter how close it is to completion. This is a clear drawback of the method. Incorporating \textcolor{black}{proper} information on the reaction progress into the resetting protocol will thus surely lead to larger accelerations. 
In this work, we develop a method to accelerate MD simulations through informed stochastic resetting (ISR). The key difference of this approach is that simulations are only reset if certain predefined criteria are fulfilled, e.g., when the distance from the target along a chosen collective variable (CV) is greater than some threshold. 
ISR is part of a broader class of spatially dependent resetting protocols~\cite{Roldan2017, Falcao_2017, Tucci2020, Plata2020, DeBruyne2020, Ali_2022, mori2023optimalswitchingstrategiesnavigation, DeBruyne2023, Bressloff_2023, Tal-Friedman2024, Keidar2024}, which were not applied before in the context of molecular simulations.

Below, we first show that ISR can accelerate MD and MetaD simulations leading to speedups that are greater than the standard resetting protocol, even for suboptimal CVs. \textcolor{black}{For a model system,} this approach leads to approximately 3.5 times greater speedups in comparison to standard SR. This translates to a speedup of  56 and 697 over unbiased simulations when used as a standalone method and combined with MetaD, respectively. We also demonstrate that ISR can lead to accelerations even in cases where standard SR fails.

The beauty of standard SR is that it provides a simple but rigorous framework to determine whether resetting will accelerate a random process and by how much.\cite{Pal2022,Reuveni2016,Evans2020, Chechkin2018}  
Recently, we generalized these results for adaptive resetting protocols that include informed resetting~\cite{Keidar2024}. 
In the second part of this paper, we use this method to compare several informed resetting criteria and determine the one that leads to the highest accelerations at almost no added computational cost. 

\textcolor{black}{In the last part of the manuscript, we show that the same procedure for inferring the unbiased mean first-passage times (MFPT) from simulations with standard resetting also works for ISR with a similar accuracy. Finally, we show that ISR is especially useful for enhanced sampling of direct transit times (DTTs)\textcolor{black}{, whose study has recently drawn experimental, theoretical, and computational interest~\cite{Makarov2021}.}
We demonstrate this on a molecular example, chignolin in explicit water, showing that ISR leads to high speedups, with a minimal compromise in accuracy.}

\section{Results and Discussion}

\subsection{Accelerating Simulations through ISR}

\begin{figure}[t]
     \centering
     \includegraphics[width=1\linewidth]{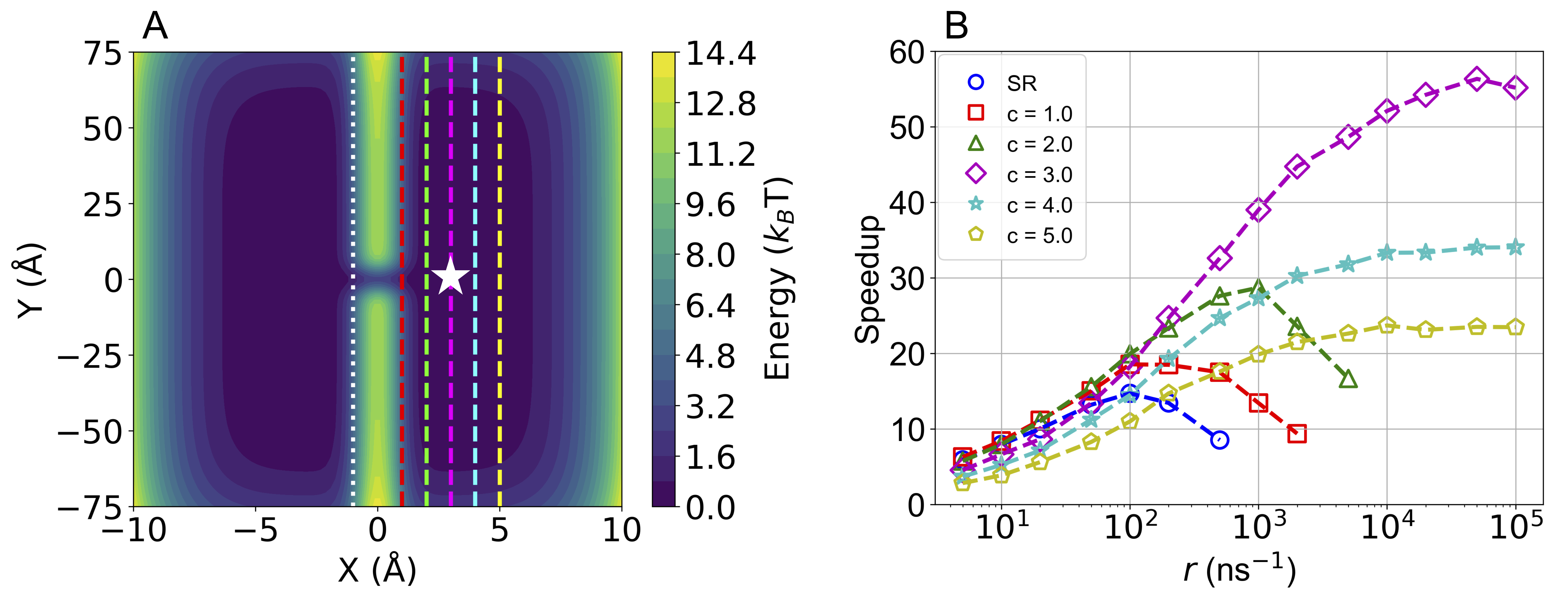}
    \caption{A) Modified Faradjian-Elber Potential, where the white star indicates the initial position, the white dotted line indicates the target where the simulations were stopped, and colored dashed lines indicate different resetting thresholds. B) The speedups obtained from standard SR and ISR with different resetting thresholds. The symbols show simulation results and dashed lines are plotted as a guide to the eye.}
 \label{fig:1}
 \end{figure}

We begin by showing that ISR can accelerate MD simulations and lead to greater speedups than standard resetting. To this end, we use a modified version of the Faradjian-Elber potential\cite{Faradjian2004}, which was previously used in ref~\cite{Blumer2024}. It is a two-dimensional symmetric well with minima located at $x =  \pm 3$ \si{\angstrom}, as shown in Figure \ref{fig:1}A. The minima are separated by a Gaussian barrier at $x=0$ \si{\angstrom}. The barrier is $12 \, k_BT$ for most 
$y$ values, but has a narrow saddle, only $3 \, k_BT$ high, around $y = 0$ \si{\angstrom}. 
All trajectories were propagated from the minimum at ($x = 3$ \si{\angstrom}, $y = 0$ \si{\angstrom}) until crossing the barrier and reaching $x < -1$ (white dashed line) \si{\angstrom}. 
Full details of the potential and simulation setup are given in the Methods section. 

We first ran $10^4$ unbiased MD trajectories with no resetting and obtained a mean first-passage time (MFPT) of $\sim 7.6$ ns. Then, we preformed simulations using standard SR, and ISR with different criteria. Resetting times were sampled from an exponential distribution with a rate $r$ ranging from 5 \(\text{ns}^{-1}\) to \(10^{5}\) \(\text{ns}^{-1}\). For standard SR, we restarted the simulations every resetting time, while for ISR, we restarted the simulations only if the $x$-coordinate value was larger than a predefined threshold $c$ at the resetting time. If this condition was not fulfilled, we continued the simulation until the next resetting time, and rechecked it. In both cases, we restart the particle to the same initial position and resample only the momentum (see Methods). We tested different values of the threshold, $c=1,2,3,4,5$ \si{\angstrom} , which are shown as vertical dashed lines in Figure \ref{fig:1}A.   The resulting speedups, defined as the ratio of the MFPT between unbiased simulations and simulations with resetting (standard or informed), are plotted in Figure \ref{fig:1}B as a function of the resetting rate. 

For standard SR (blue circles), the speedup increased with the resetting rate, reaching a maximum speedup of 15 at $r = 100$ \(\text{ns}^{-1}\). As expected~\cite{Blumer2022}, when we increased the rate further, the resulting speedup decreased \textcolor{black}{since at the infinite rate limit the particle \textcolor{black}{can} not reach the target.}
Introducing information into the resetting protocol, we obtained higher maximum speedups for all the thresholds tested. 
In this system, it is straightforward to guess that the optimal threshold, leading to the highest maximal speedup, is $c=3$  \si{\angstrom}. \textcolor{black}{The reason is that the resetting region includes all points which are further away from the target than the initial position along the $x$ direction.} In this case (magenta diamonds),  the speedup increased monotonously with the resetting rate, reaching a plateau as $r \to \infty$, which is qualitatively different than SR. We observed a similar trend for $c = 4,5$  \si{\angstrom} (cyan stars, yellow pentagons). 
On the other hand, thresholds that were closer to the target than the initial position along the $x$-coordinate, $c = 1,2$  \si{\angstrom} (red squares, green triangles), showed the same qualitative dependence of the speedup on $r$ as standard SR. 

Despite the different qualitative behavior, we stress that all thresholds led to higher maximal speedups than standard SR. Using $c = 3$  \si{\angstrom} yielded the largest speedup with a value of 56 while $c = 4,5$  \si{\angstrom} resulted in a speedup of 34 and 24, respectively. Using $c=1,2$  \si{\angstrom} resulted in speedupsof 19 and 29, respectively. 
We can explain the different behaviors of each class of thresholds by considering the effectiveness of resetting. In the case where the threshold is further away from the target than the initial position \textcolor{black}{along the CV,} resetting always helps, bringing the particles closer to the target which lowers the MFPT. We then anticipate a monotonous behavior of the speedup with the resetting rate. At the limit of $r \to \infty$, the threshold serves as a portal, teleporting every particle reaching it directly to the initial position. On the other hand, for standard resetting, or ISR with thresholds closer to the target than the initial position, resetting too frequently will prevent the particle from getting to the target. 
The difference between the two classes of thresholds is also reflected in the mean number of resetting events, as shown in Supplementary Figure 1. 
%More generally, this is true for any threshold that includes the initial position. 

We tested the sensitivity of the results to the initial position, shown in Supplementary Figure 2. We obtained a similar qualitative behavior, with higher speedups the closer the initial position is to the target. 
To further highlight the strength of this new approach, we tested it on the symmetric double-well, which is a system not enhanced by standard SR. Even in this case, ISR led to moderate speedups as shown in Supplementary Figure 3. 

\subsection{Metadynamics with Informed Resetting}

Next we studied the impact of combining ISR with MetaD to observe how the acceleration in the MFPT compares with 1) ISR alone, 2) MetaD alone, and 3) MetaD with standard SR~\cite{Blumer2024}.
We again used the Modified Faradjian-Elber potential as a model system. MetaD requires knowledge of the collective-variable (CV), which ideally describes the slowest mode of the process\cite{Demuynck2018,Peters2017}. Employing a sub-optimal CV when using MetaD often leads to reduced speedups and the inability to accurately infer the unbiased kinetics\cite{Barducci2011,Invernizzi2019,Salvalaglio2014,Besker2012,Ray2022}. 
For the modified Faradjian-Elber potential, the optimal CV is simply the $x$-coordinate (see the committor analysis in the SI). However, for most systems, identifying a good CV is a challenge, despite recent progress~\cite{peters2006obtaining,mendels_folding_2018,mendels_collective_2018,bonati_deep_2021,sidky_machine_2020,chen_collective_2021,liu_graphvampnets_2023,doi:10.1021/acs.jctc.3c00051}. 

\begin{figure}[t]
     \centering     \includegraphics[width=0.5\linewidth]{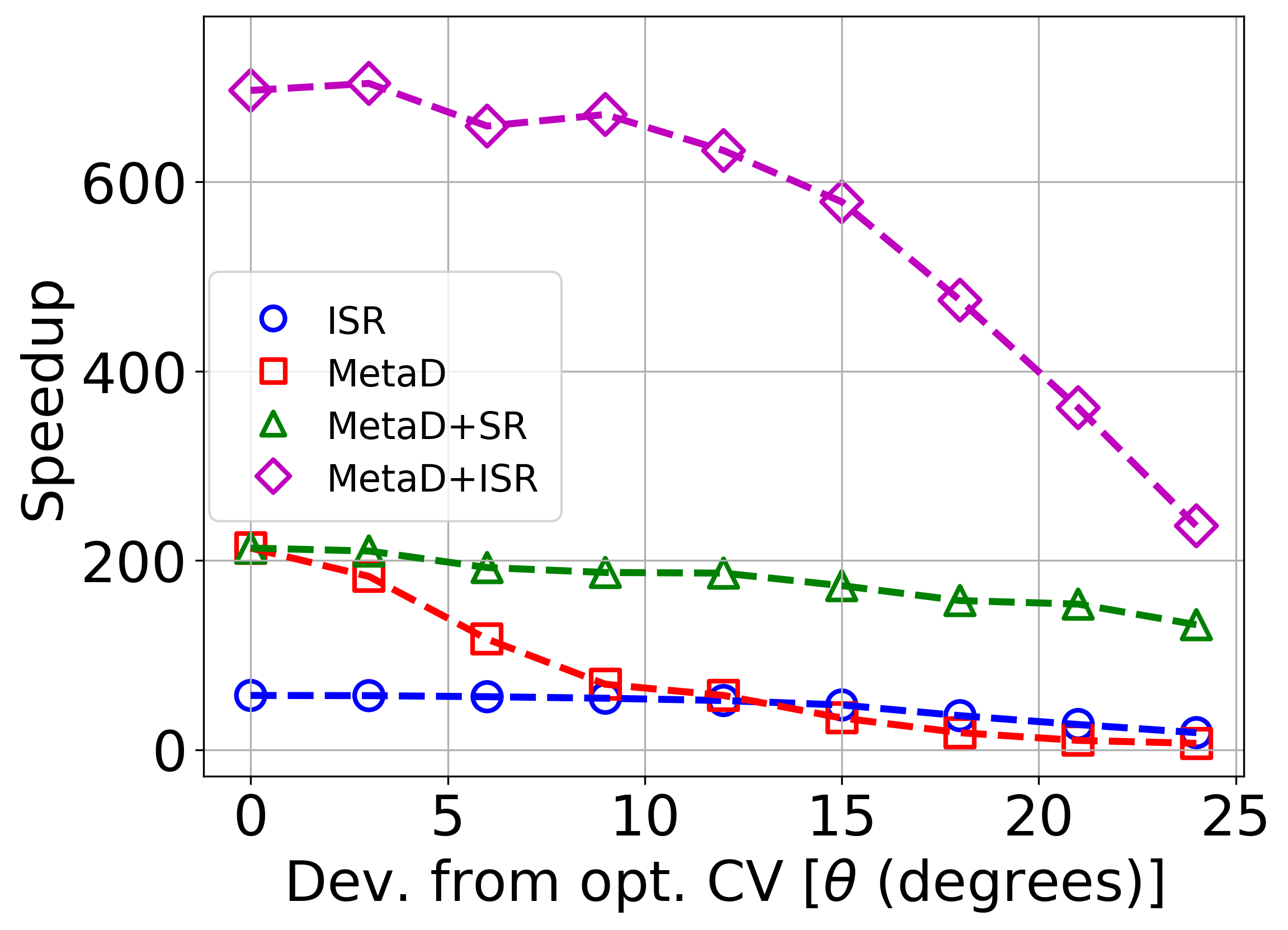}
 \caption{The maximal speedup as a function of CV quality.  We compare ISR (blue circles), MetaD (red squares), MetaD+SR (green triangles) and MetaD+ISR (magenta diamonds) using the optimal resetting rate for each CV.}
     \label{fig:2}
 \end{figure}

Here, we artificially degrade the quality of the CV by  rotating the optimal CV incrementally up to 24 degrees relative to the $x$-axis, to test the sensitivity of ISR acceleration to the CV quality. For the optimal CV, in simulations combining MetaD with ISR, the threshold was $c=3$ \si{\angstrom}. For suboptimal CVs, we rotated the threshold with respect to the initial position by the same angle as the CV.
For all CVs, and all simulations with resetting (informed or standard), we plot the maximum speedup obtained at the optimal resetting rate, see Figure \ref{fig:2}.  For all simulations with MetaD, we used a bias deposition pace of 100 time steps. 

Remarkably, we found that combining ISR with MetaD led to the highest accelerations for all CVs tested. In fact, even with the worst CV, the speedup obtained from MetaD+ISR was higher than the speedups obtained from all other methods using the optimal CV. Specifically, for the optimal CV, MetaD+ISR led to a speedup of $\sim 700$ while MetaD and MetaD+SR led to an acceleration by a factor of $\sim 200$. On the other extreme, for the worst CV, MetaD+ISR led to a speedup of $\sim 200$  while MetaD+SR led to a speedup of 132, and MetaD alone produced almost no acceleration. 

To summarize, these results show that the acceleration in the MFPT obtained from MetaD+ISR is much larger than MetaD and MetaD+SR. This is the case even for suboptimal CVs and resetting thresholds.  Similarly to standard SR, informed resetting can easily be integrated into modern MD packages to enhance simulations of chemical systems where the optimal CV is often unknown or difficult to compute. However, there remains an open question about choosing an appropriate threshold for resetting as well as determining the an efficient resetting rate, which we now address. 

\subsection{Predicting Useful Rates and Thresholds}
\label{sec:predictionsd}
So far, we performed an ensemble of simulations at \textit{every resetting rate and threshold }to assess the performance of ISR under different conditions.
However, for ISR to be a computationally viable strategy, we should be able to predict useful resetting rates and thresholds (i.e., that lead to significant accelerations) from a small set of simulations without resetting.
One of the advantages of standard SR is that very little prior knowledge is required to predict its behavior. The mean and variance of the FPTs without resetting provide sufficient information to determine if resetting will decrease the MFPT~\cite{Pal2022}. Furthermore, a small sample of unbiased trajectories $(\sim 100$), showing a single first-passage event each, gives good estimations of the optimal resetting rate and the expected speedup.~\cite{Blumer2022} This is particularly useful in combination with MetaD, where trajectories without SR are easier to sample, compared to the unbiased ensemble~\cite{Blumer2024}. Until recently, similar tools were unavailable for informed resetting. 

In a recent paper,  we developed a numerical method to obtain the MFPT for any adaptive, i.e., state- and time-dependent, resetting protocol from trajectories without resetting~\cite{Keidar2024}. We now describe how this approach can be applied to ISR with a rate $r$ and a threshold $c$. Our strategy is to formally decompose the MFPT with ISR into two contributions that can be evaluated using a set of trajectories that are sampled without resetting. For the decomposition, consider trajectories with informed resetting. Each trajectory $i$ is composed of $M_i$ segments. The first $j=1,...,M_i - 1$ segments of duration $t_i^j$ end in resetting and the final segment of duration $t_i^f$ ends in a first-passage event. Then, the overall FPT of trajectory $i$, denoted by $\tau_{i,r}$, is 
\begin{equation}
\tau_{i,r} = t_i^f +  \sum_{j=1}^{M_i -1} t_i^j =  t_i^f + \left(M_i -1\right) \, \bar{t_i} \, ,
\end{equation}
where $\bar{t_i}$ is the mean duration of segments ending in resetting for trajectory $i$. The MFPT can be then written as,

\begin{equation}
    \langle \tau \rangle_r = \langle t^f \rangle_r + \left( \langle M \rangle_r -1 \right) \langle \bar{t} \rangle_r,
    \label{eq:prediction}
\end{equation}
where we used the brackets to denote the ensemble average over the trajectories. Note that $\langle t^f \rangle_r$ can be also understood as the mean duration of a segment under the condition that it ended in a first-passage event while $\langle \bar{t} \rangle_r$ is the mean duration of a segment under the condition that it ended in resetting~\cite{Keidar2024}.
The key result of \textcolor{black}{Keidar, Blumer} et al.~\cite{Keidar2024} is that each of these terms can be easily evaluated from sampled trajectories with no resetting, as we now explain.

To do so, we run a set of $i=1,...,N$ trajectories with\textit{ no resetting}, wait until a first passage event occurs in all of them, and consider what would have happened to them under resetting. Each trajectory is of length $n_i$ steps, and has a first-passage time of $\tau_i = n_i \Delta t$, where $\Delta t$ is the simulation time step. In ISR with a rate $r$ and a threshold $c$, the probability of resetting, given that the system is in position $\boldsymbol{X}$ is given by

\begin{equation}\label{eq: probability to reset}
p(\boldsymbol{X}) = r \Delta t \cdot \begin{cases} 1 & \text{if $\boldsymbol{X} > c$} \\ 0 & \text{otherwise.} \end{cases}
\end{equation}
Then, for every trajectory $i$, we evaluate the probability that it would have survived up to time step $k$ without resetting
\begin{equation}
    \Psi_i (k) = \prod_{j=1}^{k-1} \left[ 1- p(\boldsymbol{X}_i^j) \right],
    \label{eq:survival}
\end{equation}
where $\boldsymbol{X}_i^j$ is the position of trajectory $i$ at time step $j$. We then estimate the probability that a random trajectory will survive until a first passage event with no resetting as the ensemble average, 
\begin{equation}
    \langle \Psi \rangle \approx \frac{1}{N} \sum_{i=1}^N \Psi_i(n_i).
    \label{eq:prob}
\end{equation}
Thus, with resetting, we would have had to sample $\langle \Psi \rangle^{-1}$ trajectories, on average, before observing a first passage event. We therefore estimate the mean number of segments in a simulation with resetting as
\begin{equation}
    \langle M\rangle_r=\frac{1}{\langle\Psi\rangle}\approx \frac{1}{\frac{1}{N} \sum_{i=1}^N \Psi_i(n_i)}.
\end{equation}
Next, we estimate the mean duration of the final segment in simulations with ISR as the MFPT of the trajectories without resetting, reweighed by their survival probability under resetting
\begin{equation}
    \langle t^f \rangle_r = \frac{ \langle \tau \Psi \rangle}{\langle \Psi \rangle} \approx \frac{ \sum_{i=1}^N \tau_i\Psi_i(n_i)}{\sum_{i=1}^N \Psi_i(n_i)}.
    %\frac{1}{Pr(t\leq \rho)} \frac{1}{N} \sum_{i=1}^N \tau^i \psi^i(M^i).
    \label{eq:lastsegment}
\end{equation}
Similarly, to estimate what would have been the mean duration of a segment ending in resetting, we first observe that the probability that a specific trajectory $i$ would have been reset at time step $j$ is $\Psi_i (j) p(\boldsymbol{X}_i^j)$. Averaging the segment length, $j \Delta t$, over all time steps, we get
\begin{equation}
    \bar{R_i}=\frac{1}{1-\Psi_i(n_i)}\sum_{j=1}^{n_i-1}\Psi_i(j)p(\boldsymbol{X}_i^j)j\Delta t.
\end{equation}
The normalization, $1-\Psi_i(n_i)$, is just the probability that resetting would have occurred before first passage in trajectory $i$. Finally, we estimate the average duration of segments that would have ended in resetting as the ensemble average of $\bar{R}_i$, reweighed by the probability that resetting would have occurred before first-passage, 
\begin{equation}
    \langle \bar{t}\rangle_r=\frac{\langle(1-\Psi)\bar{R}\rangle}{\langle1-\Psi\rangle}\approx\frac{\sum_{i=1}^N\sum_{j=1}^{n_i-1}\Psi_i(j)p(\boldsymbol{X}_i^j)j\Delta t}{\sum_{i=1}^N \left( 1 - \Psi_i(n_i) \right)}.
    \label{eq:resetseg}
\end{equation}
With this procedure at hand, and given a set of trajectories with no resetting, we can evaluate the MFPT at any resetting rate and threshold through Equation~\ref{eq:prediction} without running additional simulations. In practice, it can be implemented in Python with only a few lines of code, requiring a negligible fraction of the time it would have taken to actually run simulations with ISR. An example of the implementation is provided in the GitHub repository of the paper.

 \begin{figure}[t!]
     \centering
     \includegraphics[width=0.5\linewidth]{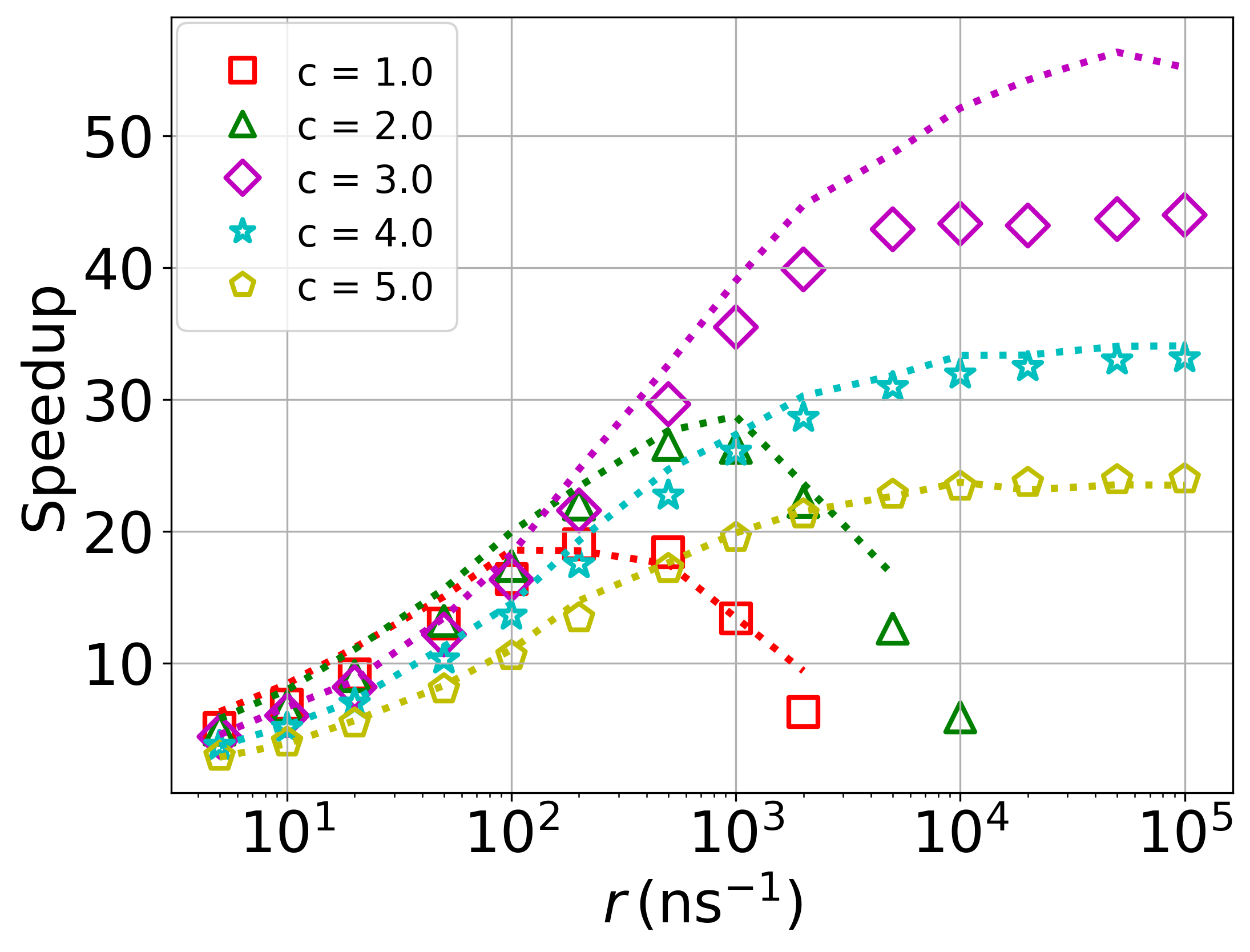} \caption{
Predictions of the speedup using different thresholds and resetting rates, for the modified Faradjian-Elber potential. Dotted lines show data obtained from simulations and presented in Fig. \ref{fig:1}B, while symbols indicate predictions from a single set of $10^4$ trajectories without resetting.
 }
    \label{fig:pre}
 \end{figure}

We first benchmark our approach by reproducing the results of Figure~\ref{fig:1}B using $10^4$ trajectories without resetting. We obtain an excellent agreement between our prediction and simulation results (see Figure~\ref{fig:pre}) for all values of $r$ and $c$, except for $c=3$ at high rates, where sampling is harder, and more trajectories will improve the predictions.
However, as mentioned earlier, for ISR to be a useful tool, we should be able to identify thresholds and rates that lead to substantial accelerations from a small set of simulations without resetting.
The use case we envision is when a small set of trajectories without resetting (standard MD or MetaD), showing a single first-passage event each, is already available. Then, screening multiple thresholds and resetting rates, to evaluate whether ISR should be used to sample more trajectories, and how efficient it will be, has practically no additional computational cost. To demonstrate this, we used only 100 trajectories with no resetting to make the predictions and identify an efficient threshold and rate. Our approach predicts high accelerations (speedup > $20$) for a threshold of $c=4$ and a resetting rate of $1000 \, \text{ns }^ {-1}$, which translates to $\sim 50\%$  of the maximum speedup shown in Figure~\ref{fig:1}.

\subsection{Kinetics Inference}

So far, we showed that ISR can accelerate molecular simulations and that we can easily identify useful thresholds and rates with a small set of simulations without resetting. However, a main goal of enhanced sampling is to be able to infer the \textcolor{black}{reaction rate, i.e., the inverse of the MFPT,} without resetting, which is hard to sample directly, from the accelerated simulations.
We next demonstrate how to infer the MFPT without resetting from simulations with ISR at a single resetting rate. The same approach that we developed for standard SR\cite{Blumer2022} can also be employed here: Using a set of ISR trajectories at resetting rate $r^*$, that led to sufficient acceleration, we predict $\langle \tau \rangle_r$ at several rates $r > r^*$ in the vicinity of $r=r^*$. Using these predictions, we approximate $\langle \tau \rangle_r$ as a function of $r$ around $r=r^*$ and extrapolate to $r = 0$ to estimate the MFPT without resetting.  Crucially, evaluating $\langle \tau \rangle_r$ at $r>r^*$ does not require performing any additional simulations, as we show in the SI. 

\begin{figure}[t!]
     \centering
     \includegraphics[width=0.5\linewidth]{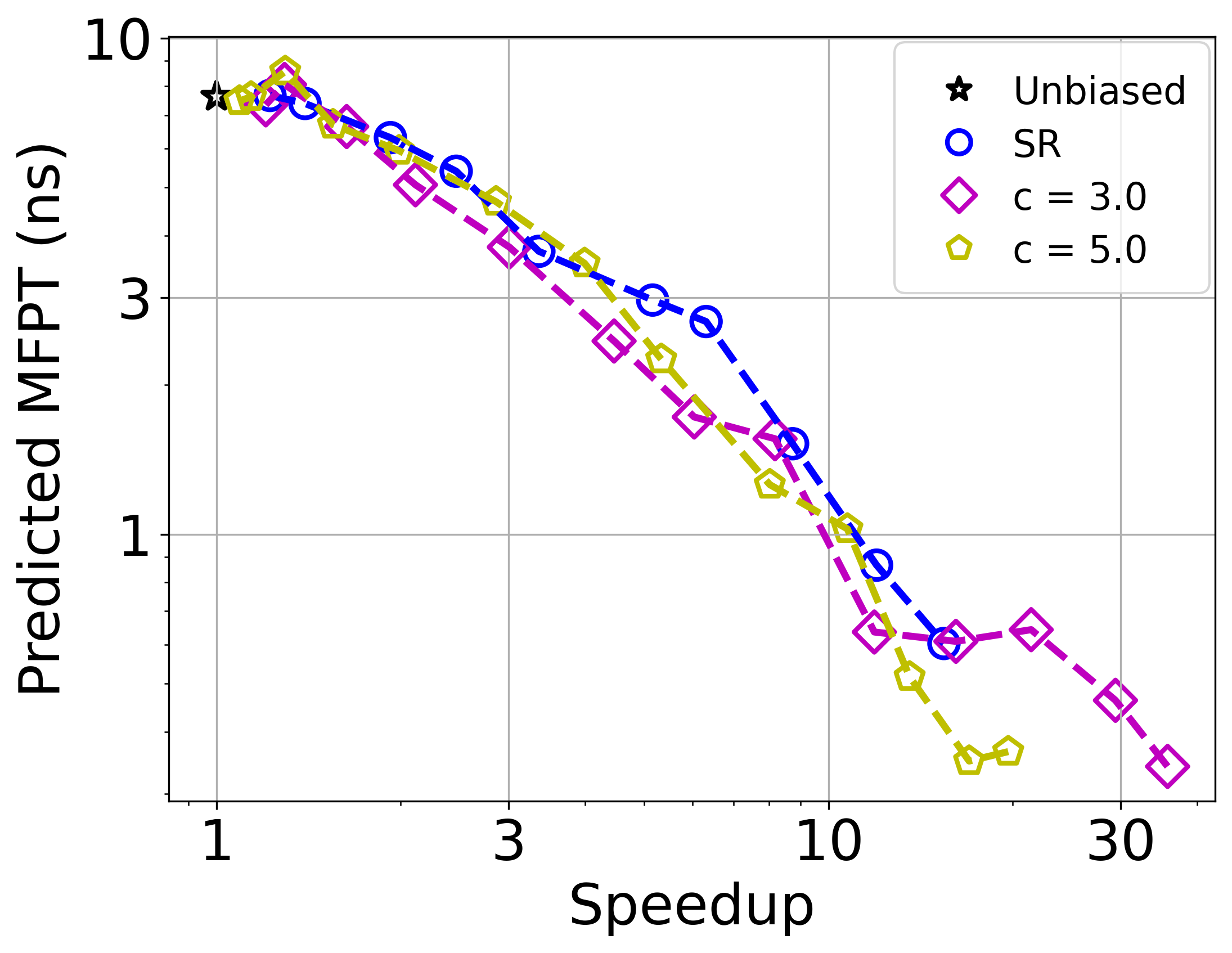}
 \caption{Predicted MFPT as a function of speedup, using SR or ISR with two different conditions. The black star gives the unbiased MFPT.}
     \label{fig:preMFPT}
 \end{figure}

Figure \ref{fig:preMFPT} shows the predicted MFPT as a function of the speedup for standard SR, and ISR with two thresholds, at different $r^*$ in the range $0.1$ to $1000 \, \text{ns}^{-1}$. A set of $1000$ trajectories were sampled for each value of $r^*$. As for other inference methods~\cite{Tiwary2013,Blumer2022,palacio-rodriguez_transition_2022,doi:10.1021/acs.jctc.4c00170}, we observe a tradeoff between speedup and accuracy. Nevertheless, we can get substantial accelerations with ISR and predict the MFPT without resetting within an order of magnitude. In comparison to standard SR,  we obtain similar accuracy at slower speedups, but can use ISR to reach higher accelerations. 
 
\subsection{Direct Transit Times in Chignolin}

\textcolor{black}{So far, in this paper and previous work on resetting for enhanced sampling~\cite{Blumer2022,Blumer2024}, we focused on kinetics inference of the unbiased MFPT from resetting-accelerated simulations. We close this paper with a new application of resetting to the inference of another important kinetic observable, the direct transit time (DTT) \cite{Elber_Makrov_Orland_Book, Makarov2021}, for which ISR is especially beneficial. We will consider a molecular example,  the mini-protein chignolin in an explicit solvent. } 

\begin{figure}[t!]
     \centering
     \includegraphics[width=\linewidth]{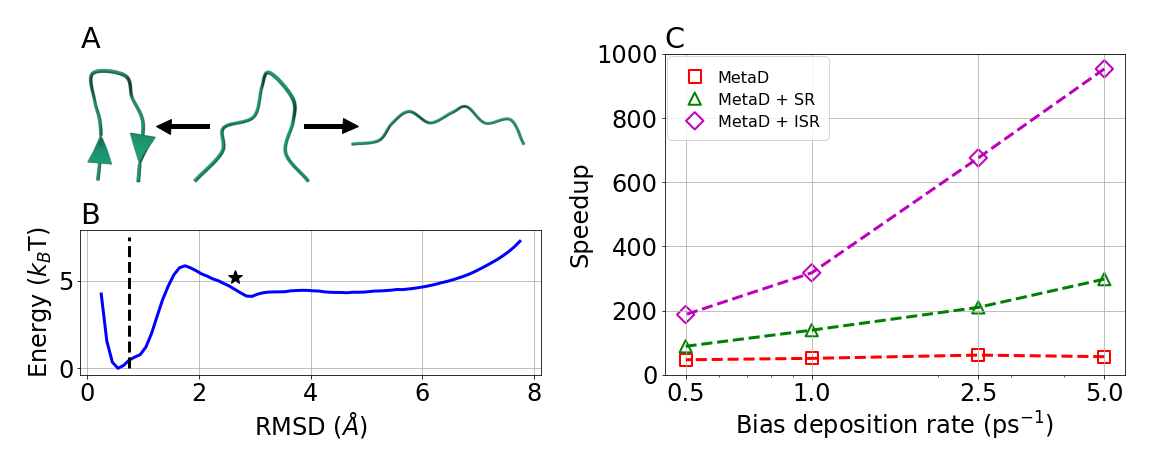}
 \caption{\textcolor{black}{A) Cartoons of representative configurations of the folded, misfolded, and unfolded states of chignolin (from left to right).
 B) FES of chignolin along the RMSD-based CV. The black star and dashed line mark the initial configuration and the FPT criterion, respectively. C) Speedup as a function of bias deposition rate for chignolin, using either MetaD (red), MetaD with SR (green), or MetaD with ISR (magenta).}}
     \label{fig:chignolin}
 \end{figure}

\textcolor{black}{Chignolin is a fast-folding protein that has a metastable misfolded state~\cite{kuhrova2012force}. Starting from it, the system can either go to the native folded state, or unfold with almost no energetic cost (cartoons of representative configurations of the states are given in Figure \ref{fig:chignolin}A). We will focus on the transition from the misfolded to the folded state. Typically, before a transition from the misfolded to the folded state occurs, several misfolding-unfolding events could happen. This is a prime example of when ISR is beneficial. It can prevent getting lost in the unfolded region of phase space, and accelerate sampling, by resetting only unfolded configurations to the misfolded state.}

\textcolor{black}{Figure \ref{fig:chignolin}B shows the FES along a CV based on the C-alpha root-mean-square deviation from a folded configuration (RMSD). The MFPT from the misfolded state (black star) to the folded state (RMSD < $0.75$ \si{\angstrom}, black dashed line) is $337 \pm 34 \, \text{ns}$, based on 100 unbiased trajectories. 
The estimated COV is 1.01, and standard SR does not provide substantial acceleration. The MFPT with ISR, on the other hand, is shorter than the unbiased one by $\sim 15 \, \%$ using no resetting at RMSD < $3.2$ \si{\angstrom} and a resetting rate of $10 \, \text{ns}^{-1}$ otherwise. This proves that ISR can accelerate molecular simulations even as a standalone method. Much larger accelerations are obtained by combining ISR with MetaD.}

\textcolor{black}{We performed 1000 MetaD simulations at several bias deposition rates, employing the RMSD-based CV. This is a suboptimal \textcolor{black}{and} naive CV, which we deliberately chose to demonstrate the power of ISR when prior knowledge is limited.
For each bias deposition rate, we estimated the optimal resetting rate for standard SR as described in Ref.~\cite{Blumer2024}, and performed simulations with standard SR at these rates.
For ISR, we followed the procedure described in a previous section, estimating the speedup as a function of resetting rate and threshold along the RMSD-based CV from the MetaD trajectories with no resetting. We then performed simulations using the rate and thresholds predicted to give the greatest accelerations.}

\textcolor{black}{Figure \ref{fig:chignolin}C compares the speedups obtained with the different methods: MetaD as a standalone tool (red), or in combination with SR (green) or ISR (magenta). MetaD provides speedups of up to a factor of $\sim 60$ without resetting. Introducing standard SR on top of it provides speedups in the range of $90-300$, while ISR provides speedups in the range of $185-950$. We find that ISR gives speedups up to 3.2 times greater than standard SR, similar to the results of the Modified Faradjian-Elber model.}

\textcolor{black}{At such high speedups, inferring the unbiased MFPT would probably result in a serious compromise in accuracy, so using ISR for this purpose is not particularly advantageous. We thus focus on a different kinetic observable, the DTT, and show that error in its inference is much less sensitive to the overall speedup.} \textcolor{black}{In a $d$-dimensional system, the DTT is the time elapsed between the last crossing of one predefined $d-1$-dimensional surface, to the first crossing of a different predefined $d-1$-dimensional surface \cite{Elber_Makrov_Orland_Book}. A common choice for these surfaces is two different values of an order parameter, centered around the transition state.} %\st{In the literature, DTTs are often also called transition path times \cite{Makarov2021}, however, to avoid confusion with transition path theory, we will not use this term. DTTs received much attention recently, since experimental measurements were done for the folding and binding reactions of biological macromolecules \cite{SungChung2012, Neupane2016, Sturzenegger2018}, stirring a lot of theoretical \cite{Satija2020, Laleman2017} and computational work \cite{Mori2016, Satija2017}.} 
\textcolor{black}{In the literature, DTTs are often referred to as transition path times \cite{Makarov2021}; however, to avoid confusion with transition path sampling~\cite{bolhuis2002transition}, we will refrain from using this term. DTTs have attracted significant interest recently due to experimental measurements of folding and binding reactions in biological macromolecules \cite{SungChung2012, Neupane2016, Sturzenegger2018}, which have spurred theoretical \cite{Satija2020, Laleman2017} and computational investigations \cite{Mori2016, Satija2017}.} \textcolor{black}{Other enhanced sampling approaches, such as Weighted-Ensemble,~\cite{HUBER199697,zhang_weighted_2010} Forward Flux Sampling,~\cite{allen2009forward} Transition Path Sampling (TPS),~\cite{bolhuis2002transition} and Milestoning,~\cite{Faradjian2004, Elber2020} could also be applied, in principle, to estimate DTTs. However, to our knowledge, most of them have not been used for that purpose. The only exception is milestoning, which has been used to estimate DTTs,~\cite{DTT_milestone} but not in explicit solvents. Since the key goal of this paper is to improve on MetaD and MetaD with standard SR as an enhanced sampling approach, by using ISR, we compare with them as a natural benchmark.}

\begin{figure}[t!]
     \centering
     \includegraphics[width=\linewidth]{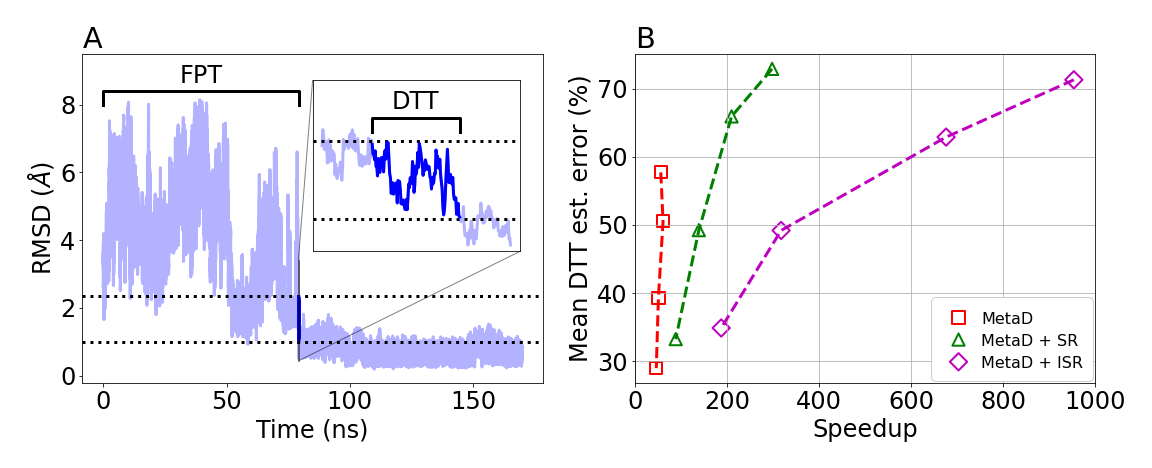}
 \caption{\textcolor{black}{A) The progress of an unbiased trajectory along the RMSD-based CV over time. The dotted lines define the DTT, and the direct transit path is highlighted in blue. The inset zooms in on a period of 250 ps.
 The black continuous lines in the main and inset axes mark the FPT and DTT, respectively.
 B) The error in the estimated mean DTT as a function of speedup using either MetaD (red), MetaD with SR (green), or MetaD with ISR (magenta).}}
     \label{fig:DDT}
 \end{figure}

\textcolor{black}{Since DTTs are much shorter than FPTs, and depend only logarithmically on the barrier height \cite{Zhang2007, Elber_Makrov_Orland_Book} we hypothesized that they will be less sensitive to the overall acceleration.} \textcolor{black}{To show this, we used enhanced sampling simulations to infer the DTTs of the chignolin misfolded to folded transition in explicit solvent. 
We defined the transition region as the RMSD interval between $1$ \si{\angstrom} to $2.35$ \si{\angstrom} (dotted lines in Figure \ref{fig:DDT}A). Direct transit paths are defined as the segment of a trajectory between the last time step with RMSD > $2.35$ \si{\angstrom}, to the first time step with RMSD $< 1$ \si{\angstrom}. To clarify, Figure \ref{fig:DDT}A shows an unbiased trajectory where the direct transit path is highlighted in blue. We see that the system remains in the unfolded regime (large fluctuations and large RMSD values) for about 80 ns, then undergoes a rapid transition to the folded state (small fluctuations around a value < $1$ \si{\angstrom}). The inset zooms in on a period of 250 ps around the direct transit path, which is 115 ps long. The unbiased mean DTT is $150 \pm 17$ ps, orders of magnitude shorter than the MFPT.}

\textcolor{black}{We estimate the unbiased mean DTT from biased simulations by simply taking the average over their DTT values. Figure \ref{fig:DDT}B shows the error in the estimations for MetaD simulations without resetting (red), with standard SR (green), and with ISR (magenta).
We find that all methods provide good estimates of the unbiased mean DTT, within much less than one order of magnitude from the unbiased value. In this case, the advantage of ISR is clear: it leads to an order of magnitude more speedup without compromising on the accuracy in the inferred mean DTT.}

\section{Conclusion}

In this manuscript, we present ISR, a new type of resetting protocol to accelerate MD and MetaD simulations. By including information about the reaction progress, e.g. by resetting only if the distance from the target along some CV is larger than a threshold, we obtain larger accelerations in the MFPT than in standard SR.
The true power of the method is, that with a \textit{single} set of $\sim 100$ trajectories, we can assess the MFPT for \textit{any} threshold and resetting rate at a negligible computational cost, and with no additional simulations. 
This enables us to predict, quickly and knowledgeably, an efficient threshold and resetting rate that can be used to significantly speed up simulations. 

\textcolor{black}{We can infer the MFPT without resetting, which is hard to directly sample, from simulations with ISR. Using the same trajectories, we can also infer kinetic observables that are less sensitive to the FPT speedup, such as the DTT, at very high accelerations without a significant loss in accuracy.}
Finally, our method is not limited to threshold ISR criteria and can be employed to any condition on any CV by adjusting Equation~\ref{eq: probability to reset} to the relevant probability~\cite{Keidar2024}. 

\section{Methods}

\subsection{The Modified Faradjian-Elber Potential}

For the Modified Faradjian-Elber potential, we employed Equation \ref{MFE} 
\noindent  \hfill 
\begin{flalign}
V(x,y)=A_{1}(x^{6}+y'^{6})+A_{2} \left(\frac{-x^{2}}{\sigma_{1}^{2}}\right) \left[1-B\exp \left(\frac{-y'^{2}}{\sigma_{2}^{2}}\right) \right],
\label{MFE}
\end{flalign}
\noindent with the following parameters:  $A_{1} = 1.2 \times 10^{-5} , A_{2} = 12, B = 0.75, \sigma_{1} = 1, \sigma_{2} = 0.5 $ and $y' = 0.1y$.

\subsection{Simulation Details}

We generated the initial conditions of each system from fixed positions with initial velocities sampled from a thermal Boltzmann distribution at $T = 300$ K \textcolor{black}{and $T = 340$ K for the modified Faradjian-Elber potential and chignolin, respectively}. We determined the first passage times of each ensemble of trajectories from the time it took the particle to pass a certain criterion which defined the target. In order to stop a trajectory upon reaching the target, we used the \textit{committor} command in the PLUMED 2.7.1 program~\cite{Bonomi2009,Bonomi2019,Tribello2014}. \textcolor{black}{For chignolin, the condition was checked only once every 1 ps.} In this work, we employed an exponential distribution of resetting times. When MetaD was combined with SR and ISR, we zeroed the bias potential after each resetting event. Every ensemble was comprised of $10^{3}$ individual trajectories, with the exception of the unbiased trajectories on the Modified Faradjian-Potential where $10^{4}$ trajectories were used in determining the MFPT\textcolor{black}{, and the unbiased/unbiased+ISR simulations of chignolin, where we collected $100$ trajectories}. 

We treated the simulations with the NVT ensemble and Langevin thermostat with a friction coefficient of \(\gamma=0.01\) \(\textrm{fs}^{-1}\). The trajectories were propagated using a time step of 1 $\textrm{fs}$. Each trajectory was comprised of a single atom with a mass of 40 g mol$^{-1}$, representing an argon atom. During MetaD, we used a bias factor of 10, along with a bias height of 0.5 \(k_{B}T\) and a grid spacing of 0.01 \si{\angstrom}. Additionally we set the Gaussian width to \(\sigma=\) 0.15 \si{\angstrom}.  

\textcolor{black}{Simulations of chignolin were performed in GROMACS 2019.6~\cite{abraham_gromacs_2015}, with the same settings as in Ref.~\cite{Blumer2024,doi:10.1021/acs.jctc.4c00170}, but a different initial configuration. To obtain this configuration, we first performed a $0.5$ ns long simulation with a strong harmonic bias around the local minima along the RMSD-based CV and a CV based on harmonic linear discriminant analysis~\cite{mendels_folding_2018}. We then performed energy minimization and ran a 0.5 ns long unbiased simulation for equilibration. For ISR, we used a customized version of the \textit{committor} command in PLUMED, which is available at the github repository.}

\begin{acknowledgement}

BH thanks Prof. Giovanni Ciccotti for the first discussion that sparked our interest in informed resetting protocols for enhanced sampling. 
BH acknowledges support by the Israel Science Foundation (grants No. 1037/22 and 1312/22) and the Pazy Foundation of the IAEC-UPBC (grant No. 415-2023).
This project has received funding from the European Research Council (ERC) under the European Union’s Horizon 2020 research and innovation program (Grant agreement No. 947731 to SR).
\end{acknowledgement}

\begin{suppinfo}

Raw data for figures and example input files are available in the GitHub repository: \\
\url{https://github.com/OfirBlumer/informedResetting}

\end{suppinfo}

\bibliography{My_Collection.bib}

\begin{tocentry}
    \centering
    \includegraphics[width=\linewidth]{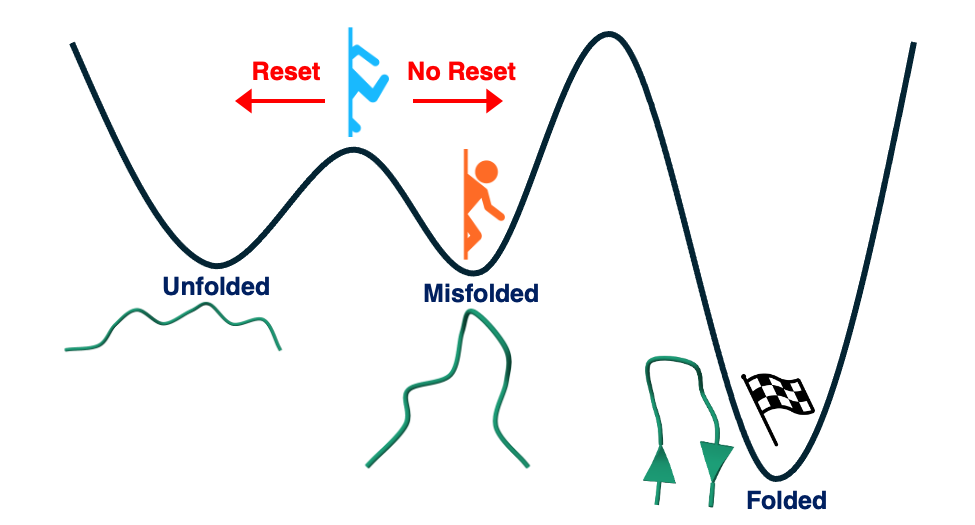}
\end{tocentry}

\end{document}

\subsection{The Mean Number of Resetting Events}
\begin{figure}[t]
    \centering
     \includegraphics[width=0.5\linewidth]{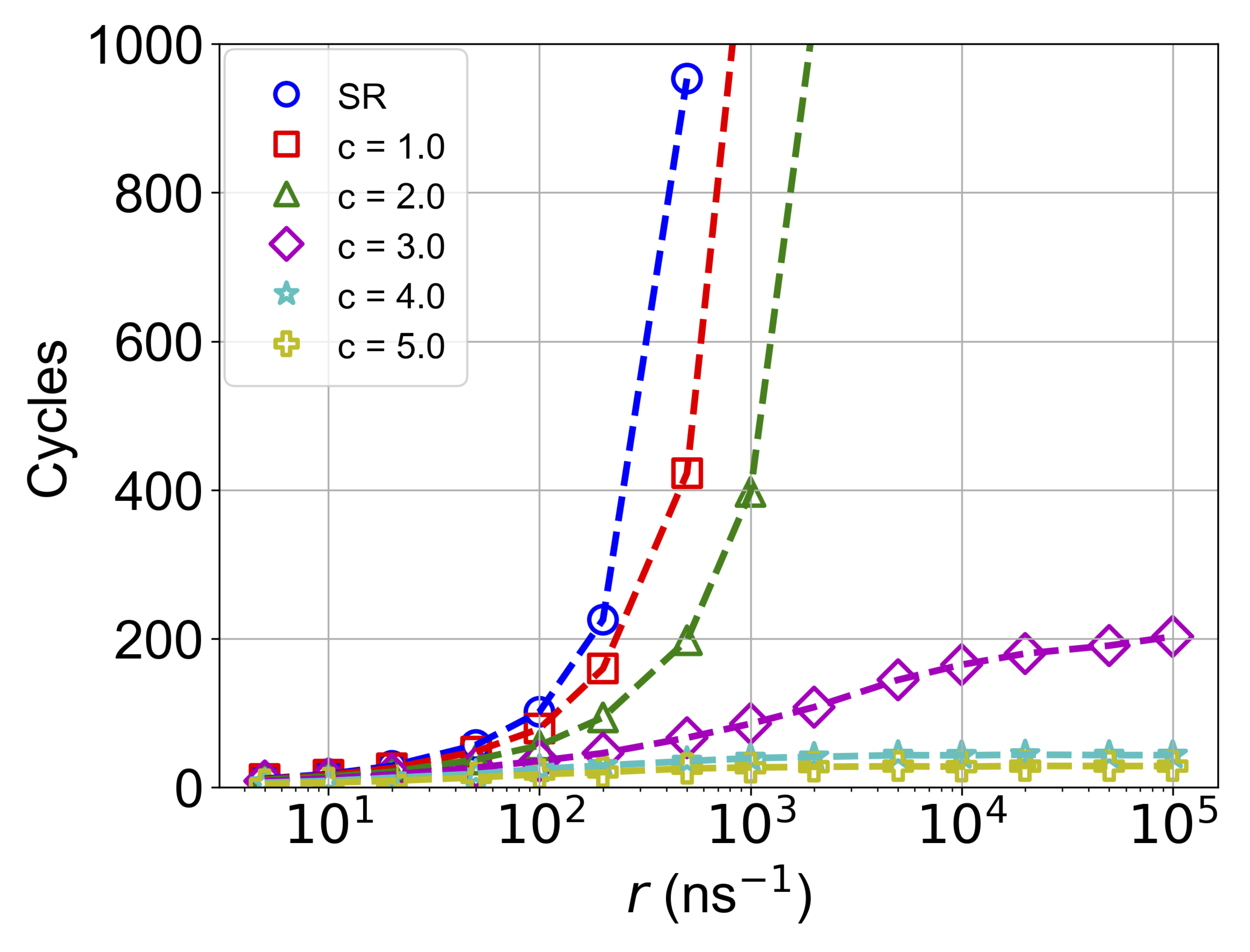}
    \caption{The mean number of resetting events when using ISR and SR for resetting rates ranging from 5 to \(10^{5}\) ns$^{-1}$.}
     \label{fig:3}
 \end{figure}

We analyzed the mean number of restarting events for the simulations of the modified Faradjian-Elber potential of Figure 1A of the main text. The results are shown in Figure \ref{fig:3}. We found that the number of average resetting events for each threshold correlated with the qualitative behavior of the speedup curves given in the main text. For thresholds which exhibited a plateau at large resetting rates we found that there was a monotonous increase in the number of average resetting events with the resetting rate until reaching a plateau. On the other hand, for thresholds in which the speedup increased, reached a maximum value, and then decreased we found that the mean number of resetting events continued to increase as the resetting rate was increased.  

\subsection{The Effect of the Initial Position}

\begin{figure}
     \centering
     \includegraphics[width=1\linewidth]{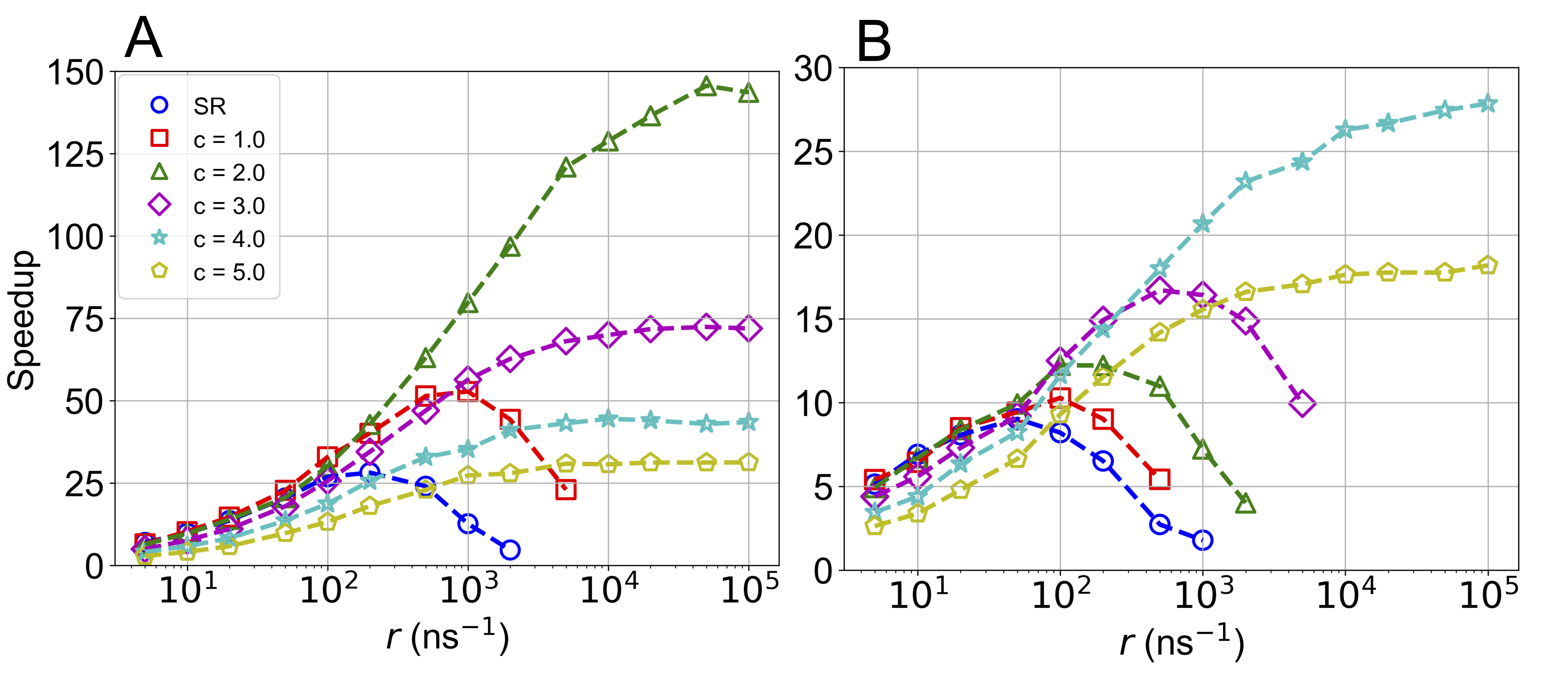}
     \caption{Speedups obtained when changing the starting location of the particle using standard SR and ISR: A) ($x, y$) = (2, 0) \si{\angstrom}, and B) ($x, y$) = (4, 0) \si{\angstrom}. The acceleration obtained for ISR and SR were studied for resetting rates ranging from 5 to \(10^{5}\) ns$^{-1}$.}
     \label{fig:4}
 \end{figure}

We analyzed the effect of restarting from different initial positions on the resulting speedup using the Modified Faradjian-Elber potential. The speedups obtained are presented in Figure \ref{fig:4}. Here, the initial position of the particle was taken to be $x = 2, 4$ \si{\angstrom} and $y = 0$ \si{\angstrom} (Figures \ref{fig:4}A, and \ref{fig:4}B, respectively.)

We found the same qualitative behavior as in the main text for all initial positions that started with $y=0$ \si{\angstrom}, with the exception that the threshold leading to the greatest acceleration changed depending on the restarting location, as expected. 

\subsection{Symmetric Double-well Potential}

\begin{figure}[H]
     \centering
     \includegraphics[width=1\linewidth]{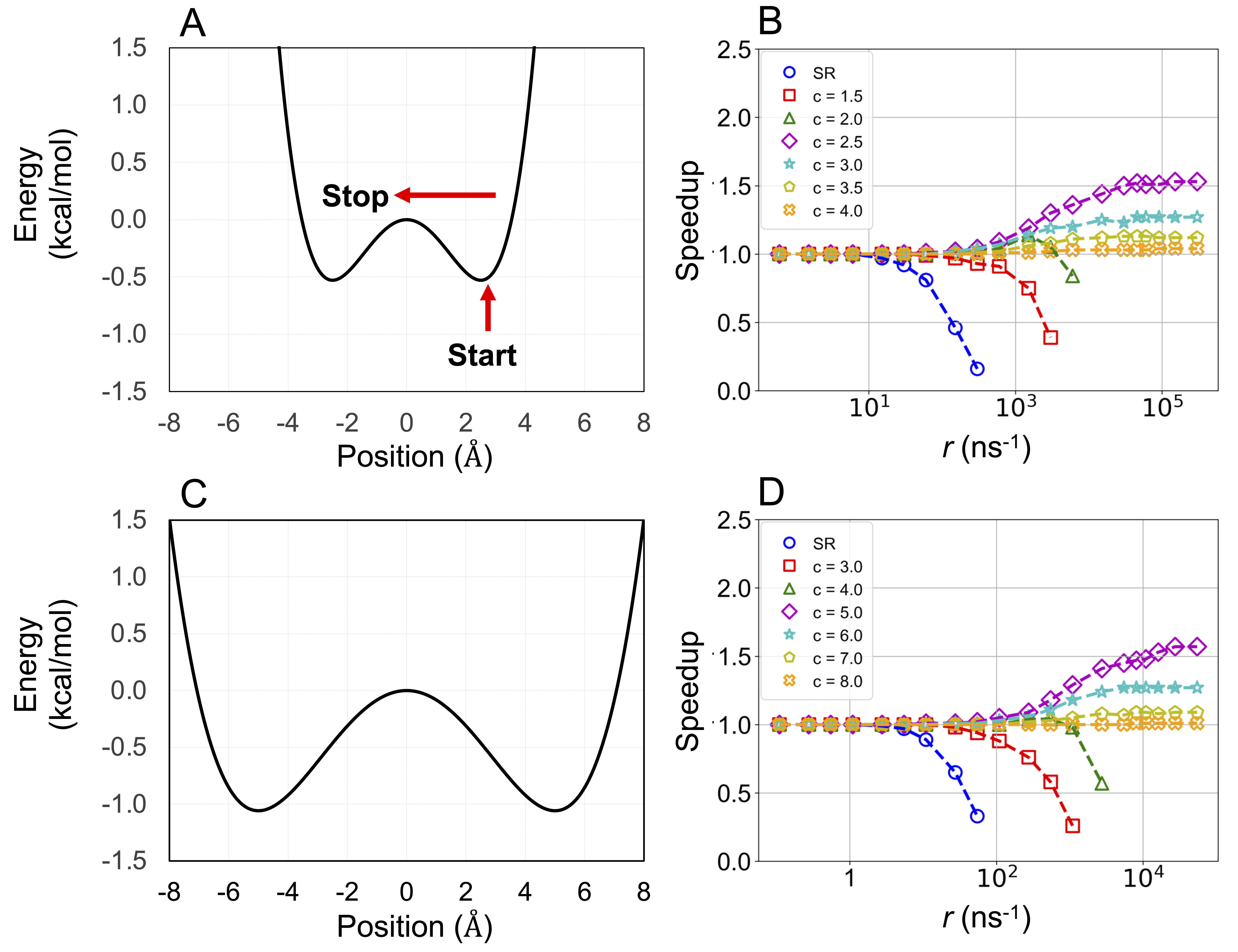}
     \caption{A) Symmetric Double Well potential with a 1.0 \(k_{B}T\) barrier and minima at ± 2.5 \si{\angstrom}, B) the speedups obtained from using SR and ISR, C) Symmetric Double Well potential with a 2.0  \(k_{B}T\) barrier and minima at ± 5 \si{\angstrom}, D) the speedups obtained from the elongated double well potential.}
    \label{fig:5}
 \end{figure}

We used the symmetric double-well potential to test if ISR can accelerate systems which do not exhibit any speedup when using standard SR (see Figure \ref{fig:5}).
The equation for the symmetric double-well potential is given by Equation \Ref{DWP},
\noindent  \hfill 
\begin{flalign}
V(x,y)=A_{1}x^{4}-A_{2}x^{2}.
\label{DWP}
\end{flalign}
\noindent Here, we modified the values of $A_{1}$ and $A_{2}$ in order to achieve the desired position of the minima and barrier height. For the potential with minima located at $\pm 2.5$ \si{\angstrom} with a 1 $k_{B}T$  barrier $A_{1}$ and $A_{2}$ were $9.404 \times 10^{17} \, \text{Jm}^{-4}$ and  $1.176 \times 10^{-1} \, \text{Jm}^{-2}$, respectively. For the elongated potential where the minima were positioned at $\pm 5$ \si{\angstrom} with a 2 \(k_{B}T\) barrier, $A_{1}$ and $A_{2}$ were  $1.176 \times 10^{17}$ $\text{Jm}^{-4}$ and  $5.878 \times 10^{-2} \, \text{Jm}^{-2}$, respectively. The trajectories were initiated from the positive $x$ minimum, $x = 2.5$ \si{\angstrom} or 5 \si{\angstrom}, and propagated until reaching the negative $x$ minimum, -2.5 \si{\angstrom} or -5.0 \si{\angstrom}. For the potential in Figure \ref{fig:5}A, the MFPT was 28.44 ps, while raising the barrier (See Figure \ref{fig:5}C) produced a MFPT of 183.60 ps.

This potential was selected because we found empirically that no acceleration could by gained by standard resetting (see ``SR" in Figures~\ref{fig:5}B and~\ref{fig:5}D). By using ISR (Figures~\ref{fig:5}B and~\ref{fig:5}D), however, we were able to obtain accelerations of up to $ \sim 50\%$. Similar to the results obtained using the Modified Faradjian-Elber Potential, the optimal threshold was associated with the initial position of the particle. 

\subsection{Committor analysis}

\begin{figure}[t]
     \centering
     \includegraphics[width=1\linewidth]{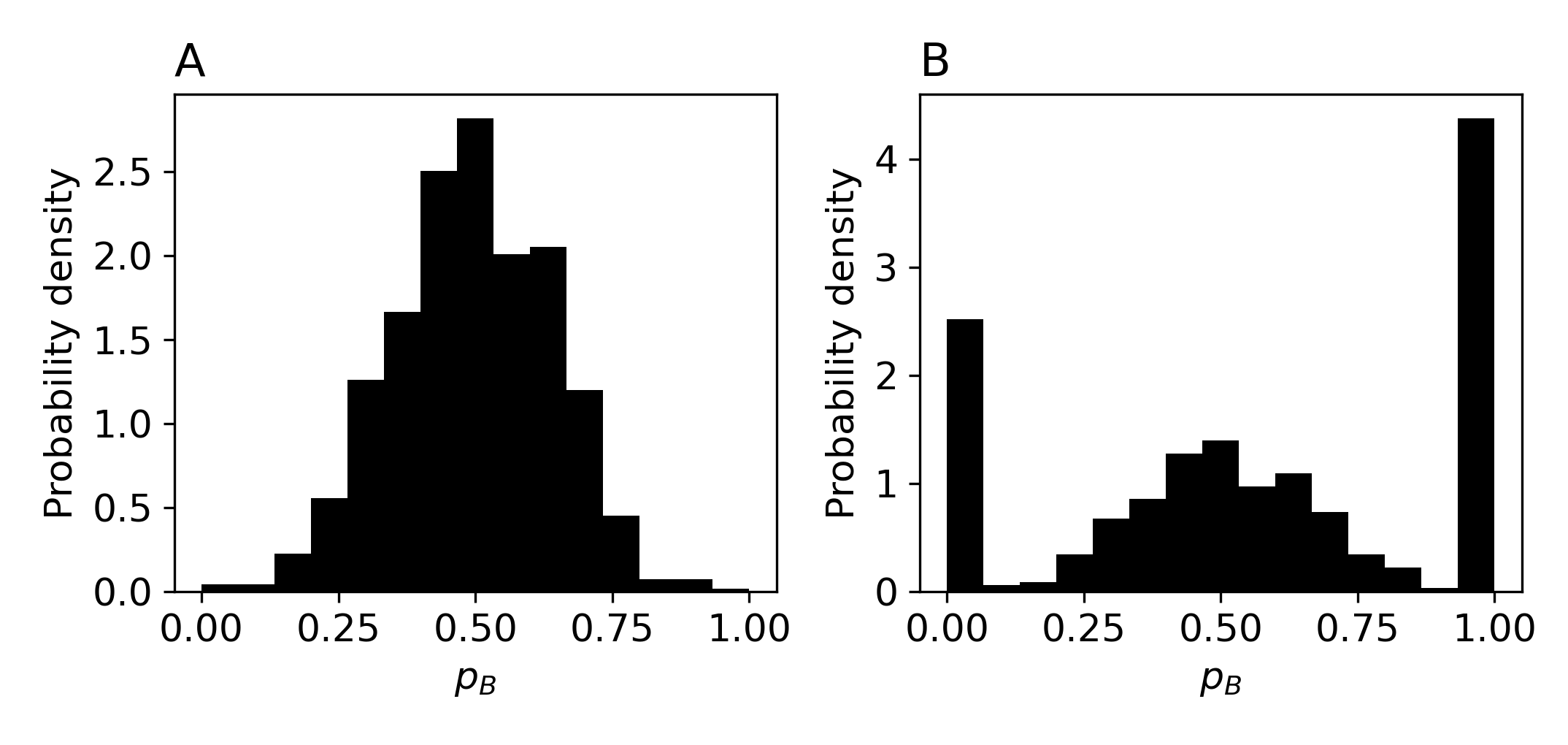}
     \caption{Histograms of $p_B$ for the modified Faradjian-Elber potential, for A) the x coordinate or B) a CV rotated $3^{\circ}$ relative to the x-axis.}
    \label{fig:6}
 \end{figure}

We conducted a committor analysis to validate our assumption that the x coordinate is the optimal CV for the modified Faradjian-Elber potential. We followed the procedure described by Peters~\cite{PETERS2017539}. We first ran a $10 \, \text{ns}$ long simulation initiated at the origin, trapped around $x = 0$  \si{\angstrom} using a trapping potential $V(x) = 500 x^2$, with $V$ being in units of $1 \, k_B T$ and $x$ in \si{\angstrom}. Then, we sampled 1000 configurations along the biased trajectory. We initialized 100 trajectories from each configuration. The trajectories where terminated when reaching either minima $x = 3$  \si{\angstrom} or $x = -3$  \si{\angstrom}, denoted states $A$ and $B$, respectively. Finally, we obtained $p_B$, the fraction of trajectories reaching state $B$ before reaching $A$, for each configuration, and plot the histogram of $p_B$ in Figure \ref{fig:6}A. The histogram is symmetric, with a peak at $p_B = 0.5$, as expected for the committor. For comparison, we performed the same procedure for a CV rotated by $3^{\circ}$ relative to the x-axis. In this case, we sampled initial configurations from two $10 \, \text{ns}$ simulations initiated at the origin. We used the same trapping potential, but also rotated by $3^{\circ}$ relative to the x-axis.
The histogram, given in Figure \ref{fig:6}B, has two peaks, at $p_B \to 0$ and $p_B \to 1$, due to an overlap between the basins and the transition state along the CV, confirming it is a suboptimal one.

\subsection{Proof that no additional simulations are required for $r>r*$}
Here, we prove that the ensemble of trajectories with ISR and resetting rate $r^*+\Delta r$ is statistically equivalent to the ensemble of trajectories obtained by performing ISR at rate $\Delta r>0$, on trajectories with ISR at rate $r^*$. 
First, notice that the resetting strategy is fully characterized by Equation 3 in the main text. So by showing that the probability of resetting is the same for two strategies, we prove that the ensembles of trajectories created by using them are equivalent. As a result, no additional simulations are required to predict the ISR results at rate $r^* + \Delta r$ for the kinetics inference. We use the standard prediction procedure, only treating the trajectories with resetting at a rate $r^*$ as the underlying process and predicting their behavior under a resetting rate $\Delta r$.

Consider first a process undergoing ISR at rate $r^*$. According to Equation 3 in the main text, the probability of undergoing resetting is
\begin{equation}
    p_{r^*}(\boldsymbol{X})=\begin{cases}
        r^* \Delta t& \text{if $\boldsymbol{X}>c$}\\
        0& \text{otherwise}.
    \end{cases}
\end{equation}

Adding ISR at rate $\Delta r$ on top, resetting at positions $\boldsymbol{X}>c$ could happen via two independent resetting mechanisms. The original one, which occurs with probability $r^* \Delta t$ and the newly added one with probability $\Delta r \Delta t$. The total probability for resetting is their sum

\begin{equation}
    p_{r^*,\Delta r}(\boldsymbol{X})=\begin{cases}
        r^* \Delta t + \Delta r \Delta t & \text{if $\boldsymbol{X}>c$}\\
        0& \text{otherwise},
    \end{cases}
\end{equation}
which is just the probability,  $p_{r^*+\Delta r}(\boldsymbol{X})$,  of resetting the original process at an ISR rate of $r^* + \Delta r$, which concludes the proof.

\bibliography{My_Collection}